# The Age of Social Sensing

Dong Wang, *Member, IEEE*, Boleslaw K. Szymanski, *Fellow, IEEE*, Tarek Abdelzaher, *Member, IEEE*, Heng Ji, *Member, IEEE*, Lance Kaplan, *Fellow, IEEE*

**Abstract**— Online social media, such as Twitter and Instagram, democratized information broadcast, allowing anyone to share information about themselves and their surroundings at an unprecedented scale. The large volume of information thus posted on these media offer a new lens into the physical world through the eyes of the social network. The exploitation of this lens to inspect aspects of world state has recently been termed *social sensing*. The power of manipulating reality via the use (or intentional misuse) of social media opened concerns with issues ranging from radicalization by terror propaganda to potential manipulation of elections in mature democracies. Many important challenges and open research questions arise in this emerging field that aims to better understand how information can be extracted from the medium and what properties characterize the extracted information and the world it represents. Addressing the above challenges requires multi-disciplinary research at the intersection of computer science and social sciences that combines cyber-physical computing, sociology, sensor networks, social networks, cognition, data mining, estimation theory, data fusion, information theory, linguistics, machine learning, behavioral economics, and possibly others. This paper surveys important directions in social sensing, identifies current research challenges, and outlines avenues for future research.

**Keywords**— Linguistic Processing, Sensor Networks, Social Networking

——————————— ◆ ———————————

## 1 INTRODUCTION

THENICALLY speaking, social sensing predates the use of physical technological sensors and social media [1]. For eons, words, either in oral or written form, have been used to communicate observed experiences to others for a variety of reasons. The messages in such communications can be both explicit and implicit. For instance, scholars may have identified secret messages embedded within the works of Plato [2]. What is new today, however, is that technology makes it much easier to formulate and share thoughts that are visible instantly to the entire world; something that took many generations for Plato.

With the greater dissemination power comes a heightened danger of misuse. It is now not only significantly easier to share ideas, but also increasingly possible to manipulate perceptions of reality at an unprecendented scale. A growing challenge today is, therefore, to find and understand the valuable and truthful messages in the much larger volume of social media content.

The unprecedented connectivity afforded by social media also gives rise to revolutionary new perspectives on empowering individuals and societies to collectively generate value from information. Pierre Levy, who introduced the concept of *collective intelligence* in the 90s, observes that human intelligence is derived from reflexive reasoning, with language being the semantic indexing scheme into the arguments and results [13]. He envisions social media to be the enabler of the next leap in reflexive collective intelligence, and describes a new meta-language, he calls the Information Economy MetaLanguage (IEML), that would empower reflexivity, facilitate discovery of semantic ties, and document information provenance (using blockchain-like technologies) to keep track of the origins of ideas and preserve the collective reasoning behind them.

If posts on a social network collectively comprise a new type of indexing into physical reality, social beliefs, concepts, biases, and ideas, the next logical question is: can one develop a new type of instrument – a new "macroscope" – to view the world state? The purpose of such a device would be to reliably observe physical and social phenomena at scale, as interpreted by the collective intelligence of social media users.

What would be the properties of such an instrument? How can these properties be influenced or optimized? How does the instrument distort the image of the world being observed? Do the posted observations themselves influence properties of the observed system, perhaps by engendering polarization and radicalization, or conversely leading to a better shared understanding, awareness, and agreement? Do they consequently affect future information propagation, thereby influencing the instrument itself? How susceptible is this instrument to intentional misuse that manipulates perceived reality, and what mitigation strategies are effective against such manipulation? Can one derive accuracy, error and performance bounds for retrieved observations, given models of human behavior and biases? Conversely, can one infer human biases and trust relations from the reported observations? An exciting aspect of these social sensing challenges is that they open up a novel research field where human-centered sciences meet research on physical, computing, and engineered systems to better understand the new instrument, and better characterize the properties of humans (posting on social media) as collective sensors.

This article focuses on one slice of the above problem; namely, the use of such a marcoscope to reconstruct *physical* (as opposed to social) reality. The benefits of addressing the above challenge are significant. Not only will better algorithms curb misrepresentation of physical reality, but also may help build smarter urban services. Most systems from urban transportation to disaster response include humans as an integral part of the underlying sensing, management, and control loops. Acting as social sensors (who post on social media), humans are able to recognize, observe, describe and report a much broader spectrum of events than do physical sensors. Especially important is humans' ability to recognize when an observation represents an abnormal pattern or behavior. Examples include calling attention to suspicious individuals or suspected crime scenes. Yet, the strength of human sensing and interpretation skills is also a liability. Unlike sensors that objectively report observed data, humans summarize observations into their own interpretations of the observed state. These interpretations are often dependent on the opinions, biases or beliefs of the observer. For example, a scene of street fighting between the police and demonstrators can be viewed by a government supporter as an example of brave police restoring law and order broken by unruly demonstrations. An alternative view could be one of a government opponent seeing it as an example of police brutality against peaceful demonstrators protesting legitimate grievances against a corrupt government. How can one map such different "semantic indexing" representations of the same event back to a neutral reconstruction of what might have actually transpired? Figure 1 summarizes the goal of building the macroscope using observations from social sensing platforms to understand the physical world in light of the cyber-physical and linguistic challenges, which are discussed next.

The above challenges can be viewed from two important deeply intertwined perspectives; namely, challenges in the cyber-physical space and challenges in the social and linguistic space.

## 2 CYBER-PHYSICAL CHALLENGES IN SOCIAL SPACES

Understanding attributes of social sensing systems requires modeling three interdependent components; (i) the humans in the loop and their cognitive models, (ii) the algorithms involved, and (iii) the laws of nature that govern the underlying physical and engineered artifacts. For example, to understand how well a disaster-response team might be able to survey post-disaster damage within a given time window will depend on (i) modeling the way survivors might respond to the disaster (including any information sharing behaviors, such as reporting of actual damage and responding to potential rumors), (ii) understanding the efficacy of software decision-support tools that distill the raw human response into more actionable information, and (iii) accounting for physical resource constraints that impact reporting and response. The confluence of these challenges requires interdisciplinary approaches to address the complex interaction between cyber, physical and social components of the holistic system. A recent Outlook issue in Computer summarized a few representative works that study related challenges of human-in-the-loop systems [3]. For example, Abowd defines the fourth generation of "collective computing" where cloud, crowd and shroud of devices connect the physical and cyber world tightly and interactively with each other. Chang describes the situation analytics, a new runtime computational model that explores user's mental states during the entire software life cycle to better understand user intention. Jiang et. al. presents their work in an emerging cyber-physical systems research area, safety-critical medical devices, and describes a state-of-the-art approach for validating closed-loop devices without jeopardizing human safety. It is informative to review the challenges brought about by their interactions when cyber-physical systems operate in social spaces.

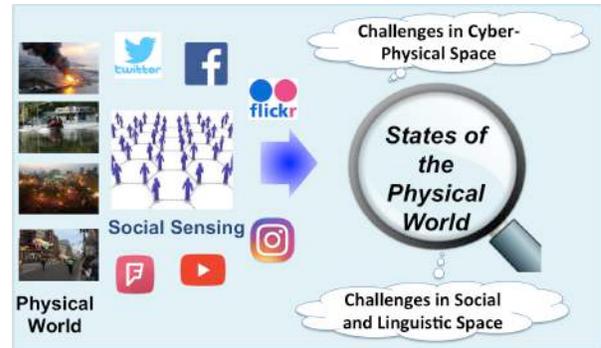

**Figure 1:** Overview of Social Sensing

Reliability is a key engineered property of cyber-physical systems. In physical sensing, measurement devices are typically well-calibrated with well-understood error properties (documented in manufacturer's data sheets). By analogy, a key social sensing challenge from a cyber-physical perspective is to understand the reliability properties of our new observational instrument; the social *macroscope* that converts social posts into an estimated state of observed phenomena. Recent history documents landmark events attributed to social media influence in the absence of information reliability assessment tools, including fake news, terrorist propaganda, and election manipulation. This generates interests in understanding signals and distortions on social media.

*Modeling instrument distortion:* One first needs to understand the nature of distortion. We can classify the distortion of information produced by human sources on social media into four broad categories. The most challenging category is intentional disinformation sent to deceive. Done purposefully (e.g., by changing small but im-portant details and then multiplying them by the power of social networks in which the malicious sources are hubs), it can cause disinformation to be considered as truth for a long time. The next category includes cases, where a person sends conclusions not sufficiently supported by data or observations, which might resonate with other people's biases and propagate further as facts. Similar, are biased interpretations by communities of people with similar opinions, distorting the recollection of events in a manner that feeds their biases, beliefs, and preferences.

Finally, there are genuine random mistakes by people processing information, such as misspelling and typos. The massive amount of sources, relative permanency of social communities involved in collecting and propagating information and our ability to partially track information provenance make it possible to build rigorous and usable formal basis for reliable truth extraction from social sensors [1].

An important consideration in social sensing is to account for the fact that humans report not events but their perception of events. These perceptions are governed by human cognition, which today can be simulated using such cognitive models as the ACT-R model [4]. Well-known limitations of human cognition, such as limited attention span, the decay of unreinforced memory traces, or limited information processing speed, impact human performance as sensors and have been adequately modeled based on ACT-R. An interesting future direction is to enable simulating crowds of agents endowed with human cognitive models to reliably account for impact of the human cognition limitations on event reporting.

The inextricable ties between information distortion, bias, cognition, and human preferences mean that, unlike the case with physical sensors, distortions on social media carry a signal themselves. These distortions are indicators of human preferences and beliefs that offer great value, for example, in tuning the design of products such as advertising campaigns, recommendation systems, and voter recruitment tools. In this article, however, we shall primarily focus on algorithms for removing the biases for a more objective representation of physical reality, where possible.

Much prior work exists on handling noisy sensors and unreliable (physical) sources. When *humans* act as sensors, however, a new model is needed to characterize the different types of distortion discussed above. From a system reliability perspective, rumors, misinformation, exaggeration, and bias propagate along social channels creating highly correlated reports (posts) and thus correlated errors at a large scale. It is well known that correlated errors decrease the reliability of systems. Social ties, trust relations, and homophily further contribute to correlated beliefs, opinions, and misrepresentations. This makes social networks a challenging instrument to use without a good understanding of their error properties and models. Recent research used simplified models of human behavior to derive fundamental information-theoretic bounds on estimation accuracy from social media [5].

*Understanding the signal:* If social media are abstracted as an instrument that measures the physical world, what is the sensing modality of this instrument? Can social sensing be treated similarly to acoustic sensing, magnetic sensing, or vibration sensing? For example, can Twitter be thought of as a medium, where a distribution of tokens of different types get emitted in response to a physical event, just the way a distribution of acoustic waves of different frequencies may be emitted in response to an acoustic target (see Figure 2)? This analogy recently led to interesting approaches for event detection, localization, and tracking, where events reported on social media are algorithmically detected in a language-agnostic manner inspired by target tracking and data fusion literature for physical signals [6]. More generally, we observe a rich set of data modalities generated by humans (e.g., text, sound, images and video). Such rich data modalities lead to interesting research questions. For example, can we explore the interdependent relationship between sensing measurements with different data modalities to obtain more accurate sensing results? How can we build an appropriate model for versatile human sensors that can simultaneously generate sensing data of diversified modalities?

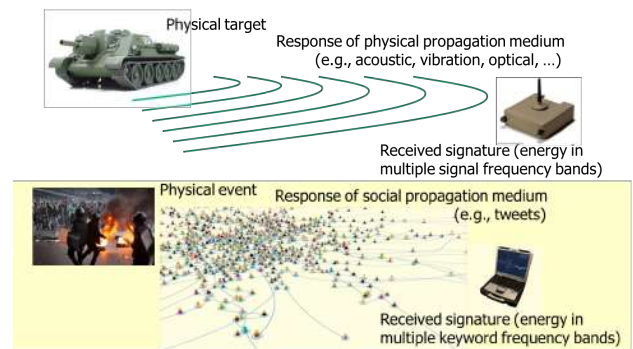

**Figure 2:** The social sensing modality; an analogy

A promising recent direction lies in developing information theoretic models of social channels. Information theory made significant headway by abstracting information as a sequence of bits. This bit stream abstraction offered ways to model noise, reason about error probabilities, derive capacity, and construct optimal decoders. Inspired by this approach, recent work on using Twitter as a data source viewed statements about physical reality as a binary signal (a bit). This is because the statement of the tweet is either true of the physical world or not, leading to the binary abstraction. Given this representation, Twitter becomes just another noisy channel, where binary signals propagate. Conceptually, the source of this channel is the physical world itself. The output constitutes human observations of the world. Distortion is introduced, for example, when people make up rumors. This is equivalent to flipping a bit from 0 (something that was *not* true of the physical world) to 1 (something claimed true at the output of the social medium). Similarly, failures to report true events can be modeled as bit omissions, and claims that deny actual events are bit flips from 1 to 0. One can then reason about the probability of different bit flips and omissions, and derive estimators of social channel input (i.e., physical world state) given its output, using existing estimation-theoretic techniques. The approach has been successful at modeling statements about objectively observable realities (e.g., "it is raining") where ground truth is unique and unambiguous. It needs extensions to cases where ground truth is undefined or subjective. For example, "John is a great president!".

*Quantifying data reliability and performance bounds:* Recent maximum-likelihood estimation techniques adopting the above binary model show much promise in jointly detecting the original state of the world from the distorted output of the social channel, together with estimating the distortion itself. This problem is known in classical data fusion as a joint signal detection and channel estimation problem, offering the necessary analytic foundations for the solution approach. An optimal estimation-theoretic framework can jointly estimate both the reliability of data items posted, and credibility of information sources involved without prior knowledge of either [1].

Estimation theory offers solid analytical foundations for bounding the error of an optimal estimator. Representing social media as noisy binary communication channels, as described above, allows estimation-theoretic frameworks to become applicable to the reliability analysis of data cleaning systems operating on outputs of social channels. A key recent contribution that leverages the aforementioned insight exploits expressions of error bounds of maximum-likelihood estimators to assess the quality of estimation results on social media [1]. This quality analysis is immensely important in practical settings, where errors have consequences. However, the analysis is complicated by two factors. First, it needs to account for correlated errors that result from rumor-spreading behaviors, when a person reports as their own observations received from others without verification. Second, the analysis needs to account for the expressed degree of vagueness in human observations, such as "the protest is *possibly* unsafe." In such cases, the estimator must take into account the degree of confidence that a source expresses in its messages to make proper assessments.    Recent work based on subjective logic, a type of uncertain probabilistic logic, has developed a framework to assess source reliability in such situations [7]. This framework presumes that the expressions of vagueness are quantified as specific probabilities. The conversion of natural language expressions of vagueness to quantifiable numbers is a very difficult problem that remains to be fully solved (see Section 3).

*Exploring the role of dependencies between sources:* Dependent sources are common in social sensing applications resulting in uncertain data provenance. Namely, it is not unusual for sources to report observations they received from others as if they were his/her own. This is a common case in the applications where humans play the role of data sources that are connected through social networks (e.g., follower-followee relationship on Twitter and friends relation-ship on Facebook.). The rumor spreading behavior of human sensors has no analogy in correctly functioning physical sensors. From a cyber-physical system perspective, this means that errors in "measurements" across sources may be non-independent, as one erroneous observation may be propagated by other sources without being verified. Recent works attempt to either develop source selection schemes to carefully select independent sources on social networks or build reliable social sensing models to explicitly model the source dependency into the social signal processing engine [1]. The complex and dynamic source dependency graphs on social networks deserve more investigation.

*Understanding communities, social trust and polarization:* The influence of relations between sources on results of the social macroscope is perhaps most evidenced in observing human communities. Humans interact, operate, exchange and propagate information much more frequently within the communities than across them. Communities enable members to develop trust between each other. Thanks to either individual's selection of community to join or community member's influence on each other, communities tend to increase level of homophily among their members [8].  This often results in members having similar opinions, attitudes, and beliefs, as well as similar misconceptions, misrepresentations, and susceptibilities. Communities serve specific needs of their members, so the level of the homophily among members if often strong on traits related to these needs, like skills, access to information, or interests, such as particular political movements, specific sports, or preference for a specific genre of music [9]. Typically each person belongs to several communities, serving different needs and with varying level of community association. The notion of communities is often associated with a notion of trust and shared opinions or biases. It is therefore a key notion when it comes to understanding information reliability.

Going back to the Twitter example, the information propagation characteristics of the social channel are largely affected by trust relations among sources, as well as by their views. Individuals are much more likely to rebroadcast information they heard from a source they trust. Hence, when the same information is reported by multiple sources, the reliability of this information cannot be accurately computed without understanding the underlying social trust network. From a reliability perspective, trust relations increase correlations in reporting, and hence the probability of correlated errors among individuals. Bias towards a particular opinion also affects information propagation. Individuals tend to propagate claims they agree with. Accordingly, understanding community biases is important in assessing error correlations and consequently data reliability.

Equally interesting is the inverse problem. Since trust and biases modulate propagation of information, observing the propagation patterns themselves reveals the underlying trust relations and biases of sources [7]. This effect is demonstrated in Figure 3 showing information dissemination topologies among sources supporting different candidates in an Egyptian election. The problem is interdisciplinary. Social science informs the development of generative models that predict how individuals modulate information propagation. Given such models, it is possible to estimate the position of each individual on issues of interest by observing the collective information propagation patterns on the social network.

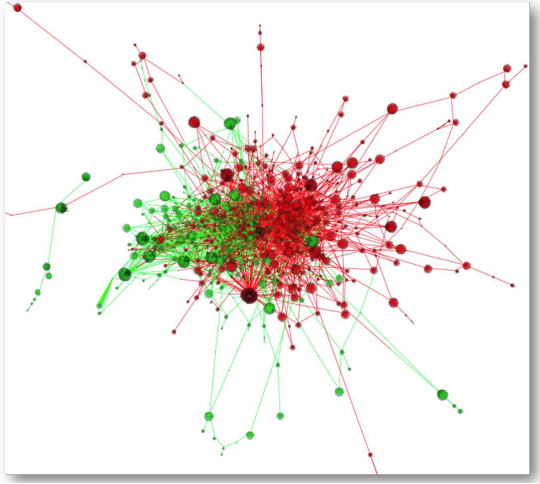

**Figure 3:** Information dissemination topologies of pro- and anti-sources.

A particularly interesting problem is to estimate ground truth from social media outputs in the presence of polarization, a situation when two (or more) distinct camps in the social network propagate largely disjoint sets of claims that are often conflicting. Recent work addressed the challenge of detecting polarization and identifying bias of individual sources in polarized communities [10]. Experiments showed that, when accounting for such bias, state reconstruction from social observations tends to align more closely with ground truth in the physical world than when polarization is not considered. It is also possible to estimate the polarity-dependent propagation networks from tweet propagation patterns, which enables computing correlations among sources for the purposes of computing their error-dependence properties and hence inform fact-finder design.

*Fusion of physical and social sensors:* With error bounds computed on social sensing observations, an interesting challenge arises to develop fusion engines that combine physical and social sensing data within the sensing-control-actuation loops of cyber-physical systems. While social networks can be viewed as additional sensing sources that corroborate physical sensors, fusion of social and physical sensing can also give rise to capabilities that are new in kind. For example, such fusion can explain anomalies seen by sensors in view of data collected from social media [11]. Interesting research questions remain to be answered. For example, how to accurately correlate and fuse data from physical and social sources? How to effectively handle different data modalities (e.g., numerical, text, images, video)? How to automatically infer the casual relationships between identified events and quantify the accuracy of such inference?

Finally, sensing systems exist in the context of control loops. Current systems typically feature loops with only a small number of well-designed sensors. Can future applications use crowd-sensing as a sensing subsystem in their control loops? How to analyze the closed loop characteristics of such systems? How to determine the impact of information errors on closed-loop stability and performance optimization? In many such systems, individuals (besides being data sources) will also be part of system "control". Decision-makers will act on the data in ways that impact the physical state being observed. Hence, an interesting question is to understand what information to present to the decision-maker (and how) in order to offer the best decision-support despite the inherent noise in the underlying social channel.

## 3 CHALLENGES IN THE LINGUISTIC SPACE

To complete our estimation of the state of the physical world, media content must be properly understood. This can be thought of as the challenge of "interfacing" to the human sensor. To address this challenge, we need to seek dramatic advances in rapid low-cost development of Information Extraction (IE) and Text Mining technologies for understanding social media content.

Today's state-of-the-art human language processing technologies either rely on (i) *supervised learning*, which suffers from the high cost of large-scale manual annotation and the limited predefined fact types, or on (ii) *unsupervised learning*, which usually yields unsatisfactory performance due to the exclusion of language/task-specific knowledge. We need to systematically discover and unify latent and expressed knowledge from traditional symbolic semantics and modern distributional semantics through advanced machine learning models. The idea of unified representations of semantics has been a focus of investigation both in social science (e.g., the seminal work of Pierre Lévy on meta-languages for expressing collective intelligence) and computing (e.g., the recent interest in language embedding as a way to abstract semantic spaces).

An information extraction system should be able to simultaneously discover a domain-rich schema and extract information units with fine-grained types. It should have a "cold-start" and be adaptable to any domain, genre, language, or data modality without any human annotated data. Recent work on combining symbolic semantic and distributional semantics [12] has made it possible to discover semantic schema and extract facts simultaneously, without relying on human defined and constructed ontologies for specific domains [13]. It should also be able to adapt within a few hours to a new scenario using very few resources. The new framework should rapidly acquire, categorize, structure and zoom in on incident-specific expectations from various non-traditional sources, including human in the loop. The output of such a framework will constitute a structured set of data types and their instances/values, derived from text (for example, to describe the state of an observed environment). Such an output will be more amenable to integration within sensor data fusion systems, than the original unstructured text generated by human observers. The process of collecting information introduces many challenges elaborated below.

*Ambiguity in a Sentence:* A core challenge in processing natural language on social media is that important information is often presented implicitly. Such presentation contains a wide amount of imprecision, ambiguity, vagueness and implicit information. Natural Language Processing (NLP) technologies currently rely heavily on surface processing. This makes it difficult to exploit deep structure, background knowledge and source information. Most discussion forum posters assume the readers already know the on-topic entities and events, and thus they don't bother to elaborate the background for these target entities. Also, they tend to use short and informal mentions for efficient discussions. As a result, the local contexts in which an entity is mentioned are often not sufficient for disambiguation

An entity linker needs to automatically construct a background knowledge base for disambiguation. Without knowing the global topic knowledge about an entity, entity linking systems tend to mistakenly link it to a more popular entity. Finally, entity disambiguation techniques are still weak on exploiting commonsense knowledge.

*The Importance of Context:* The current state-of-the-art considers informal text from social media an impoverished alternative to formal text (e.g., newswire), producing erroneous and conflicting facts from informal and noisy data. Innovative techniques on implicit and morphed information extraction are required to handle imprecise language. For example, in some societies, censorship and surveillance are imposed over information on the Internet. These actions include blocking websites, deleting posts, and filtering information about a specific entity or event. With the growth of online social networking services, people rapidly invent new ways to communicate sensitive ideas. We call this phenomenon information morphing [14], which also existed in the traditional forms of communications. Aside from the purpose of avoiding censorship, other motivations for morphing include expressing sarcasm/verbal irony or positive sentiment toward some entities or events, planning for secret actions, trading illegal products, masking politically unpopular views, or simply creating interesting conversations. Morphing raises unique challenges for entity and event co-reference resolution. Document-level profile and corpus-level temporal distribution-based features can narrow down the scope of referents, but deeper understanding is needed to discover concealed facts. While ambiguity, redundancy and implied information can pose significant problems for natural language understanding systems, the impact of these errors can be mitigated by incorporating disparate, deep sources of information.

*Discourse Ambiguity:* Besides sentence-level and sub-sentence-level ambiguities, there is yet one more level of ambiguity: super-sentential, or discourse level ambiguity which goes beyond sentence boundaries. This third level of ambiguity comes from two sources. The first is co-reference ambiguity; pronouns often refer to entities outside of the current sentence and it is ambiguous to which entity they refer. Failure to resolve these undermines our ability to understand the original discourse. The second source of ambiguity is discourse structural ambiguity; a discourse, like a sentence, has its own internal structure, often represented as a tree or graph. For deep understanding of the text, we need to know, for example, which sentence is the topic sentence, which sentence is the elaboration or contrast of another sentence, and the temporal structure between sentences (in a story line).

*Expressions of Fuzziness and Vagueness:* Human sources typically describe objects using fuzzy terms, e.g., "she is *very tall*" instead of precise terms, e.g., "she is *6'2''* tall." It is very difficult to convert fuzzy terms into numbers as the use of such terms varies wildly over different humans or societies. It is much easier to calibrate physical sensors whose noise performance can be consistently measured. The problem becomes more difficult as humans add notions of vagueness in their reports (e.g., "she is *quite* tall"). The conversion of such terms into a distribution of possible height values is very much an open and difficult problem.

## 4 FUTURE DIRECTIONS

While research on various components of the social sensing vision is currently underway, the field lacks a unified interdisciplinary problem formulation that takes a holistic approach to modeling humans as sensors, and modeling social media as noisy measuring instruments or channels. Several challenges remain. For example, how to incorporate language ambiguities discussed above into the channel bit error models considered by ground-truth estimators? How to augment information theory to account for semantic errors and approximations? When individuals summarize information, how does the resulting lossy compression affect downstream error propagation? Since sources often implicitly assume some shared background with the receiver, how to predict errors in interpretation that arise because of mismatch between the backgrounds of a message source and message recipients? How to encode inference mechanisms that help recognize corroborating evidence? How should these mechanisms identify and handle irreconcilable observations? What is the right level of abstraction at which human behavior (in relaying information) needs to be modeled? Finally, actions of individuals in regard to information interpretation and processing before propagation are often correlated to the degree of the connectivity, familiarity and trust between individuals. Hence, in addition to actions of individuals, there is a need for understanding impact of relations among individuals and influence of social groups and communities.

These questions call for the involvement of at least five synergistic disciplines. Specifically, the work needs *(i) computational social scientists* to model human behavior, at both the individual and community scale, and quantify its susceptibility to errors, omissions, deceit, and other irregularities, *(ii) linguists* to model strengths and imperfections of human communication and their compounding effects on reliability of information dissemination, *(iii) information theorists* to model social networks as imperfect communication media and derive fundamental capacity limits and uncertainty envelopes, *(iv) data mining experts* to investigate the impact of the underlying error models on reliability of knowledge extraction from imperfect information, and *(v) cyber-physical experts* to develop estimation-theoretic observability and control limits and tools that offer closed-loop robustness guarantees in the face of derived capacity limits, uncertainty envelopes, and knowledge errors.

The research challenges and future directions reviewed here call for the emergence of a new field that combines social and cognitive models, linguistics, estimation theory, information theory, and reliability analysis, with the goal of putting social media exploitation on well-understood analytic foundations, not unlike fusion of hard data from physical sensors and signals. New interdisciplinary research is outlined to bring about novel solutions for a better theoretical and systematic understanding of emerging social sensing systems in a future sensor and media-rich world.

## ACKNOWLEDGEMT

Research was sponsored by the Army Research Laboratory and was accomplished under Cooperative Agreements Number W911NF-09-2-0053, W911NF-17-2-0196 and W911NF-17-1-0409, DARPA under Award W911NF-17-C-0099, National Science Foundation (NSF) under Grants No. CBET-1637251, CNS-1566465 and IIS-1447795, and Office of Naval Research (ONR) Grant No. N00014-15-1-2640. The views and conclusions contained in this document are those of the authors and should not be interpreted as representing the official policies, either expressed or implied, of the Army Research Laboratory, DARPA, ONR, NSF, or the U.S. Government. The U.S. Government is authorized to reproduce and distribute reprints for Government purposes notwithstanding any copyright notation here on.

## AUTHOR BIOS

DONG WANG is an assistant professor at the Department of Computer Science and Engineering, the Universi-

ty of Notre Dame. He received his Ph.D. in Computer Science from University of Illinois at Urbana Champaign (UIUC) in 2012. He has authored/co-authored more than 60 referred publications in the area of big data analytics, social sensing, cyber-physical computing, real-time and embedded systems. Wang's interests lie broadly in developing analytic foundations for reliable sensing and information distillation systems, as well as the foundations of data credibility analysis, in the face of noise and conflicting observations. He is a member of IEEE and ACM. Contact him at dwang5@nd.edu.

BOLESLAW SZMANSKI is the Claire and Roland Schmitt Distinguished Professor of Computer Science and the Founding Director of the Social Cognitive Networks Academic Research Center at Rensselaer Polytechnic Institute, Troy, NY. He received his Ph.D.in Computer Science from Institute of Informatics in Warsaw Poland. He is an author/co-author of over three hundred publications and an editor of six books. Dr. Szymanski is an IEEE Fellow and the foreign member of National Academy of Sciences in Poland. His interests focus on network science, especially social networks and computer networks. Contact him at szymab@rpi.edu.

TAREK ABDELZAHER is a Professor and Willett Faculty Scholar at the Department of Computer Science, the University of Illinois at Urbana Champaign. He received his Ph.D. from the University of Michigan in 1999 on Quality of Service Adaptation in Real-Time Systems. He has authored/coauthored more than 200 refereed publications in real-time computing, distributed systems, sensor networks, and control. Abdelzaher's research interests lie broadly in understanding and influencing performance and temporal properties of networked embedded, social and software systems in the face of increasing complexity, distribution, and degree of interaction with an external physical environment. He is a member of IEEE and ACM. Contact him at zaher@cs.uiuc.edu.

HENG JI is Edward P. Hamilton Development Chair Professor in Computer Science Department of Rensselaer Polytechnic Institute. She received her Ph.D. in Computer Science from New York University. Her research interests focus on Natural Language Processing and its connections with Data Mining, Social Science and Vision. She coordinated the NIST TAC Knowledge Base Population task since 2010, and served as the Program Committee Chair of NAACL-HLT2018. She is a member of IEEE and ACM. Contact her at jih@rpi.edu.

LANCE KAPLAN is a researcher in the Networked Sensing and Fusion branch of the U.S Army Research Laboratory (ARL). He received the B.S. degree with distinction from Duke University, Durham, NC, in 1989 and the M.S. and Ph.D. degrees from the University of Southern California, Los Angeles, in 1991and 1994, respectively, all in Electrical Engineering. Dr. Kaplan serves on the Board of Governors for the IEEE Aerospace and Electronic Systems (AES) Society (2008-2013, 2018-Present) and as VP of Conferences for the International Society of Information Fusion (ISIF) (2014-Present). Previous, he served as Editor-In-Chief for the IEEE Transactions on AES (2012-2017) and on the Board of Directors of ISIF (2012-2014). He is a Fellow of IEEE and of ARL. His current research interests include information/data fusion, reasoning under uncertainty, network science, resource management and signal and image processing. Contact him at lance.m.kaplan.civ@mail.mil.